\documentclass[amsmath,prl]{revtex4}
\usepackage{amssymb,amsmath,amsthm,amsfonts,eucal,mathrsfs,amsbsy}
\usepackage{graphicx}
\usepackage{epstopdf}
\usepackage{color}
\definecolor{rojo}{RGB}{255,40,0}

\begin{document}
	
\title{The rise and fall of the amplitude, and phase, around Exceptional Points: a Scattering matrix approach\\
}
	
\newcommand{\TGb}[1]{{\color{blue} [TG] {#1}}}
\newcommand{\TGr}[1]{{\color{red} [TG] {#1}}}
	
\author{J. Col\'in-G\'alvez}
\email{colinlug05@gmail.com}
\affiliation{Departamento de F\'isica, Universidad Aut\'onoma Metropolitana-Iztapalapa,
	A. P. 55-534, 09340 Ciudad de M\'exico, Mexico.}
	
\author{E. Casta\~no}
\email{ele@xanum.uam.mx}
\affiliation{Departamento de F\'isica, Universidad Aut\'onoma Metropolitana-Iztapalapa,
	A. P. 55-534, 09340 Ciudad de M\'exico, Mexico.}
	
\author{G. B\'aez}
\email{gbaez@azc.uam.mx}
\affiliation{Departamento de Ciencias B\'asicas, Universidad Aut\'onoma Metropolitana-Azcapotzalco,	Av. San Pablo 420, Col. Nueva Rosario, 02128, Ciudad de M\'exico, Mexico.}

\author{V. Dom\'inguez-Rocha\footnote{Corresponding author}}
\email{vdr@azc.uam.mx}
\affiliation{Departamento de Ciencias B\'asicas, Universidad Aut\'onoma Metropolitana-Azcapotzalco,	Av. San Pablo 420, Col. Nueva Rosario, 02128, Ciudad de M\'exico, Mexico, and\\
Departamento de F\'isica, Universidad Aut\'onoma Metropolitana-Iztapalapa,
A. P. 55-534, 09340 Ciudad de M\'exico, Mexico.}

\begin{abstract}
We analyze the behavior of a non-Hermitian opened one-dimensional quantum system with $\mathcal{PT}$ symmetry. This system is built by a dimer, with balanced gains and losses described by a parameter $\gamma$. By varying $\gamma$ the system resonances, which are naturally separated, coalesce at the exceptional point (EP). The transmission spectrum is obtained by means of the scattering matrix ($S$ matrix) formalism and we examine the wave functions corresponding to the resonances as a function of $\gamma$. Specifically, we look for the behavior and distribution of the phases of the $S$ matrix before, at and after the EP.
\end{abstract}

\maketitle
	

A complete generalization of non-Hermitian quantum mechanics has not been reached yet since Bender proposed, more than two decades ago, a new subset of  systems in which the energy spectrum has an imaginary part~\cite{Bender1998}. The Hamiltonians of this subset, which are called quasi-Hermitian, have the peculiarity of being invariant under the application of the symmetry operators of parity (${\cal P}$) and time (${\cal T}$). Also, regardless of whether this Hamiltonian is real or complex it has real eigenvalues~\cite{Bender2002, Bender2007}. Systems with ${\cal PT}$-symmetry have a parameter associated with non-hermiticity that allows the transition between a real spectrum and a complex one. This special value was called Exceptional Point (EP) by Kato in 1966~\cite{Kato}. EPs appear when the variation of that parameter allows the coalescence of two (or more) naturally repelling eigenvalues, as well as their eigenvectors, of the associated Hamiltonian operator ~\cite{Alu2021, Heiss}. Quasi-Hermitian systems and EPs have been extensively studied in various areas of physics such as electromagnetism~\cite{electromagnetism2020}, optics~\cite{optics2017, Chong2011, Ramezani2011}, microwaves~\cite{Richter}, acoustics~\cite{acoustics2016}, photonics~\cite{photonics2019}, elasticity~\cite{vdomEP}, quantum mechanics~\cite{Schomerus, quantummechanics2018, Ortega, Dorey}, electronics~\cite{electronics2017} or atomic physics~\cite{atomicphysics2007}. EPs have been applied in improving the sensitivity of optical measurements~\cite{optics2017}, in robust transfer of energy wireless~\cite{electronics2017}, in unidirectional invisibility~\cite{Chong2010, atomicphysics2007}, slowing light~\cite{Moiseyev2018}, in alteration of spectral bands~\cite{Zhen2015, Kartashov2018}, or in accelerometers~\cite{KottosNat}.

In order to try to bring some additional information to the global discussion, in this letter we wonder how the distribution of the phases of a scattering system changes as a function of the non-hermicity parameter $\gamma$. To do this, we study the properties of an open one-dimensional non-Hermitian quantum system which presents at least one EP. The system is made up of a ${\cal PT}$-symmetric dimer, where one of the scatterers has losses while the other one has gains. Both losses and gains are balanced through $\gamma$. We study the transmission spectra by developing analytical expressions for the scattering matrix ($S$ matrix) of the dimer. We show that the transmission increases its amplitude as $\gamma$ grows in the exact ${\cal PT}$-phase but it starts decreasing its amplitude as we move further into the broken ${\cal PT}$-phase. In fact, the $S$ matrix tells to us about the unitarity of the system. On the one hand, it is quite common to think that a system has absorption if $S$ is subunitary ($R+T<1$). On the other hand, we must think that a system is emiting if $S$ is over unitary~\cite{Schomerus} ($R+T>1$). The system we are studying has both behaviours. In the process of showing this behaviour, we analyze the wave functions of the resonances involved in the EP. The phases of the $S$ matrix in the Argand plane support this behaviour by increasing its radii before and after the EP. Even more surprising is the fact that the distribution of the phases cluster around a certain angle when they were distributed all over the Argand plane.

\begin{figure}[h]
	\centering
	\includegraphics[width=8.6cm]{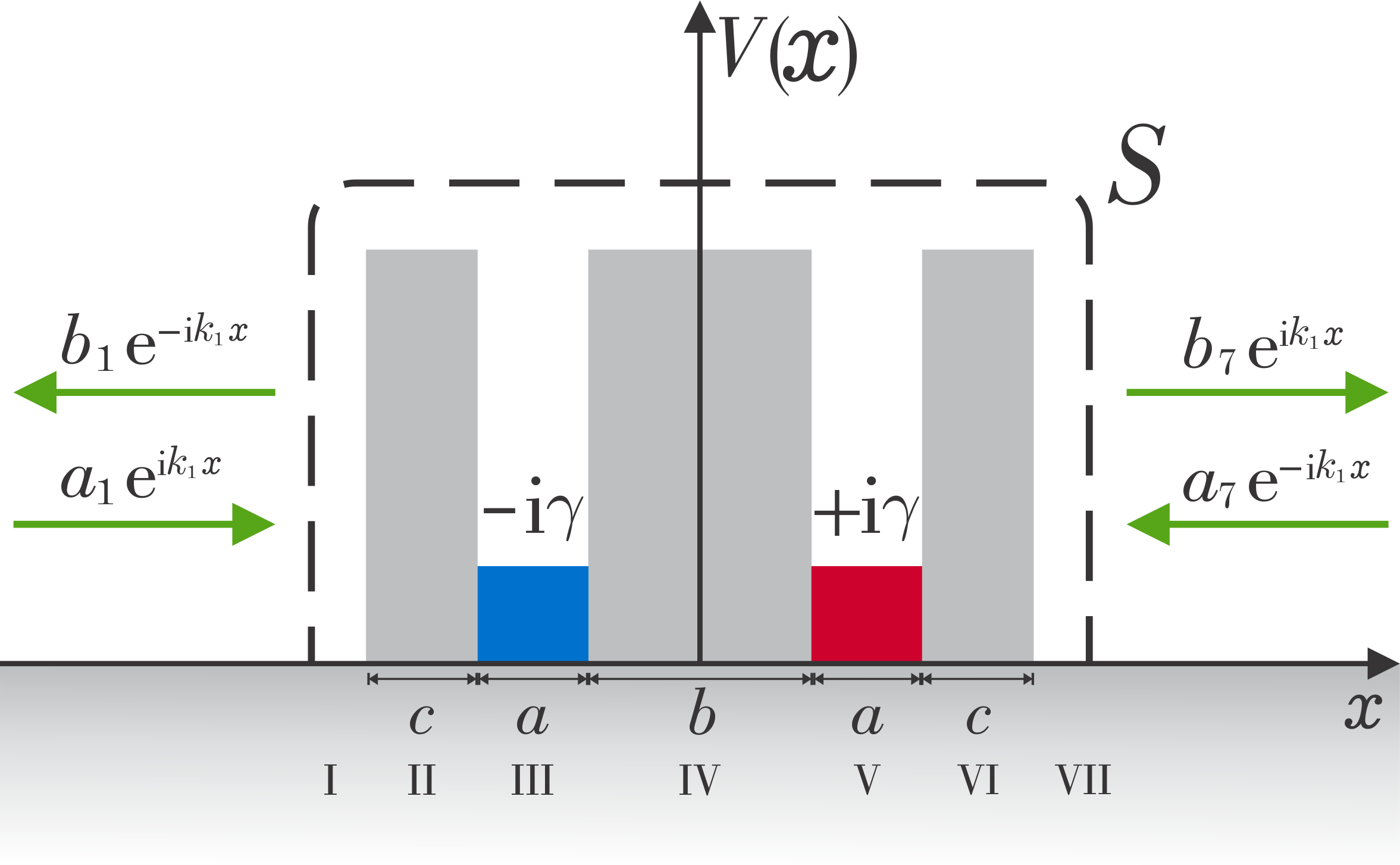}
	\caption{(Color online) Diagram of a dimer composed of two scatterers, with three potential barriers each one. The potential of the outer barriers of each scatterer is real, while the potential described by the central barrier is complex. The sign of the imaginary part of the central barrier gives rise to losses (blue region) or gains (red region). As expected, this system is quasi-Hermitian since it is invariant when applying the operators ${\cal PT}$.}
	\label{fig1:Diagram}
\end{figure}

The system we are studying is a one-dimensional open quantum dimer with balanced gains and losses, parameterized by $\gamma$, to which plane waves are incident from both sides. The dimer is made up of two scatterers whose real part of the potential that describes them is identical, while the signs of the imaginary part are opposite to each other. Each scatterer of the dimer is composed of three potential barriers as illustrated in Figure~\ref{fig1:Diagram}. The potential that describes the whole system is composed by seven regions. It is real for all regions except for regions III and V where the potential is complex. In regions I and VII the potential $V(x)=0$; $V(x)=V_\textrm{b}$ for regions II, IV and VI; while the loss (gain) potential is $V'-\textrm{i}\gamma$ ($V'+\textrm{i}\gamma$) for region III (V). The width of the region IV is the sum of the barriers of each dimer scatterer since they are contiguous. For each of these regions the stationary solutions of the  Schr\"odinger equation gives a superposition of plane waves traveling to the left and right given by $\psi_1(x)=a_1\textrm{e}^{\textrm{i}k_1x}+b_1\textrm{e}^{-\textrm{i}k_1x}$, and $\psi_i(x)=b_i\textrm{e}^{\textrm{i}k_ix}+a_i\textrm{e}^{-\textrm{i}k_ix}$ for $i=2,3,\cdots,7$. The wavenumbers are $k_1=k_7=\sqrt{2mE/\hbar^2}$, $k_2=k_4=k_6=\sqrt{2m(E-V_\textrm{b})/\hbar^2}$, $k_3=\sqrt{2m\left[E-(V'-\textrm{i}\gamma)\right]/\hbar^2}$, and $k_5=\sqrt{2m\left[E-(V'+\textrm{i}\gamma)\right]/\hbar^2}$, where $m$ is the mass of the particle, while $E$ is its energy. We are only interested in resonances, consequently the energy must be bounded between the heights of the barriers as $V_\textrm{b}>E>V'$.

We are only interested in resonances, therefore the energy must be $V_\textrm{b}>E>V'$.
We restrict ourselves to the energy region $V_\textrm{b}>E>V'$ because we are focus on the resonance phenomena
It could be interesting to study this phenomenon in the region below
We restrict ourselves to the energy region since we are not interested in the evanescent case

Resonance phenomena

The scattering matrix relates the amplitudes of the outgoing waves ($b_1$ and $b_7$) in terms of the incoming ones ($a_1$ and $a_7$)~\cite{PereyraBook, MelloBook}. The $S$ matrix is of $2\times2$ because the system has two opened channels~\cite{FerryBook}, namely
\begin{equation}
	\begin{bmatrix}
		b_1 \textrm{e}^{\textrm{i} k_{1}\left(a+\frac{b}{2}+c\right)} \\
		b_7 \textrm{e}^{\textrm{i} k_{7} \left(a+\frac{b}{2}+c\right) }
	\end{bmatrix}
	= S
	\begin{bmatrix}
		a_1 \textrm{e}^{-\textrm{i}k_{1}\left(a+\frac{b}{2}+c\right)} \\
		a_7 \textrm{e}^{-\textrm{i} k_{7} \left(a+\frac{b}{2}+c\right) }
	\end{bmatrix}\nonumber. 
\label{ec1:Sorigin}
\end{equation}
The $S$ matrix has the form
\begin{equation}
	S=\begin{bmatrix}
		r & t'\\
		t & r'    
	\end{bmatrix},
	\label{ec2:Smatrix}
\end{equation}
where $r$ and $t$ ($r'$ and $t'$) are the amplitudes of reflection and transmission of the dimer when the waves are incident from the left (right). We must emphasize that the dependence of $r, t, r'$ and $t'$ on $\gamma$ is implicit in the wavenumber. Explicitly these elements are
\begin{eqnarray}
	r&=& r_{12}                                                  
	+t'_{12}\; \frac{1}{\textrm{e}^{-2\textrm{i} 
			k_1c}-r'_{12}\;r_{23}} \; r_{23}  \;t_{12}                   
	+t'_{12}\; 
	\frac{\textrm{e}^{-\textrm{i}k_{1}c}}{e^{-2\textrm{i} 
			k_1c}-r'_{12}\;r_{23}}\;  t'_{23} 
	\;\frac{1}{\textrm{e}^{-2\textrm{i}k_{2}a}-r'_{1-3}\;r_{34}}\;r_{34}
	\;t_{23}\; \frac{\textrm{e}^{-\textrm{i} k_1 
			c}}{\textrm{e}^{-2\textrm{i} k_1c}-r'_{12}\;r_{23}} \;t_{12} \nonumber\\
	&+&  t'_{12}\frac{\textrm{e}^{-\textrm{i} 
			k_{1}c}}{\textrm{e}^{-2\textrm{i} k_1c}-r'_{12}\;r_{23}}\; t'_{23} 
	\frac{\textrm{e}^{-\textrm{i}k_{2}a}}{\textrm{e}^{-2\textrm{i} 
			k_{2}a}-r'_{1-3}\;r_{34}}
	t'_{34}  
	\frac{1}{\textrm{e}^{-2\textrm{i} k_{1}b}-r'_{1-4}\;r_{45}}\;r_{45} 
	\;  t_{34} \frac{\textrm{e}^{-\textrm{i} k_{2} 
			a}}{\textrm{e}^{-2\textrm{i} k_{2}a}-r'_{1-3}\;r_{34}}
	\;t_{23}\frac{\textrm{e}^{-\textrm{i} k_1 
			c}}{\textrm{e}^{-2\textrm{i} k_1c}-r'_{12}\;r_{23}}   
	\;t_{12}                                                     \nonumber\\
	&+&t'_{12}\;\frac{\textrm{e}^{-\textrm{i}k_{1}c}}{\textrm{e}^{-2\textrm{
				i}k_1c}-r'_{12}\;r_{23}}\;t'_{23}\;\frac{\textrm{e}^{-\textrm{i}k_{2}a}}
	{\textrm{e}^{-2\textrm{i} 
			k_{2}a}-r'_{1-3}\;r_{34}}
	\;t'_{34}\;\frac{\textrm{e}^{-\textrm{i}k_{1}b}
	}{\textrm{e}^{-2 \textrm{i}k_{1}b}-r'_{1-4}\;r_{45}} \;t'_{45}\; 
	\frac{1}{\textrm{e}^{-2\textrm{i} k_{2}a}-r'_{1-5}\;r_{56}}  \;r_{56} 
	\nonumber\\
	&\times&t_{45}\frac{\textrm{e}^{- \textrm{i}k_{1} b}}{\textrm{e}^{-2 \textrm{i}k_{1}b}-r'_{1-4}\;r_{45}}\; t_{34}\;\frac{\textrm{e}^{-\textrm{i} k_{2} 
			a}}{\textrm{e}^{-2\textrm{i} 
			k_{2}a}-r'_{1-3}\;r_{34}} 
	 \;t_{23}\;\frac{\textrm{e}^{-\textrm{i}k_1 
			c}}{\textrm{e}^{-2\textrm{i}k_1c}-r'_{12}\;r_{23}} \;t_{12}\\
	&+&t'_{12}\;\frac{\textrm{e}^{-\textrm{i} 
			k_{1}c}}{\textrm{e}^{-2\textrm{i} 
			k_1c}-r'_{12}\;r_{23}}\;t'_{23}\;\frac{\textrm{e}^{-\textrm{i}k_{2}a}}{
		\textrm{e}^{-2\textrm{i} k_{2}a}-r'_{1-3}\;r_{34}}
	\;t'_{34} 
	\;\frac{\textrm{e}^{-\textrm{i} k_{1}b}}{\textrm{e}^{-2\textrm{i} 
			k_{1}b}-r'_{1-4}\;r_{45}}\; t'_{45} \;\frac{\textrm{e}^{-\textrm{i} 
			k_{2}a}}{\textrm{e}^{-2\textrm{i} k_{2}a}-r'_{1-5}\;r_{56}}\;\nonumber\\
	&\times&t'_{56} 
	\frac{1}{\textrm{e}^{-2\textrm{i} k_{1}c}-r'_{1-6}\;r_{67}}\;r_{67}\; t_{56}\frac{\textrm{e}^{-\textrm{i} k_{2} 
			a}}{\textrm{e}^{-2\textrm{i} k_{2}a}-r'_{1-5}\;r_{56}} 
	\;t_{45} 
	\frac{\textrm{e}^{-\textrm{i} k_{1} b}}{\textrm{e}^{-2 \textrm{i} 
			k_{1}b}-r'_{1-4}\;r_{45}}\;t_{34}\frac{\textrm{e}^{-\textrm{i} k_{2} 
			a}}{\textrm{e}^{-2\textrm{i} 
			k_{2}a}-r'_{1-3}\;r_{34}}	
	\;t_{23}\frac{\textrm{e}^{-\textrm{i} k_1 
			c}}{\textrm{e}^{-2\textrm{i} k_1c}-r'_{12}\;r_{23}}\;t_{12}\nonumber,
	\label{eq2:reflection}
\end{eqnarray}
\begin{equation}
	t=t_{67}\frac{\textrm{e}^{-\textrm{i}k_{1} 
				c}}{\textrm{e}^{-2\textrm{i} k_{1}c}-r'_{1-5}r_{67}}t_{56} 
		\frac{\textrm{e}^{-\textrm{i} k_{2} a}}{\textrm{e}^{-2\textrm{i} 
				k_{2}a}-r'_{1-4}r_{56}}	
	t_{45} \frac{\textrm{e}^{- \textrm{i}k_{1} 
				b}}{\textrm{e}^{-2 
				\textrm{i}k_{1}b}-r'_{1-3}r_{45}}t_{34}\frac{\textrm{e}^{-\textrm{i} k_{2} a}}{\textrm{e}^{-2\textrm{i} k_{2}a}-r'_{1-2}r_{34}} 
				t_{23}\frac{\textrm{e}^{-\textrm{i}k_1 
				c}}{\textrm{e}^{-2\textrm{i}k_1c}-r'_{12}r_{23}}  t_{12},
	\label{ec2:transmision}
\end{equation}
\begin{eqnarray}
	r'&=&r'_{67}
	+t_{67} \; \frac{1}{e^{-2\textrm{i} k_{1}c}-r'_{56}r_{67}} \; 
	r'_{56} \; t'_{67}
	+t_{67} \; 
	\frac{\textrm{e}^{-\textrm{i}k_{1}c}}{\textrm{e}^{-2\textrm{i}k_{1}c}
		-r'_{56}r_{67}}  \; t_{56}  \; 
	\frac{1}{\textrm{e}^{-2\textrm{i}k_{2}a}-r'_{45}r_{7-5}}  \; r'_{45}  
	\; t'_{56} \;  
	\frac{\textrm{e}^{-\textrm{i}k_{1}c}}{\textrm{e}^{-2\textrm{i}k_{1}c}-r_
		{67}r'_{56}}  \;  t'_{67}\nonumber\\
	&+&t_{67}  
	\frac{\textrm{e}^{-\textrm{i}k_{1}c}}{\textrm{e}^{-2\textrm{i}k_{1}c}
		-r'_{56}r_{67}} \; t_{56} \; 
	\frac{\textrm{e}^{-\textrm{i}k_{2}a}}{\textrm{e}^{-2\textrm{i}k_{2}a}
		-r'_{45}r_{7-5}}
	\; t_{45}
	\frac{1}{\textrm{e}^{-2\textrm{i}k_{1}b}-r'_{34}r_{7-4}} \;  r'_{34} \; 
	t'_{45}   
	\frac{\textrm{e}^{-\textrm{i}k_{2}a}}{\textrm{e}^{-2\textrm{i}k_{2}a}
		-r'_{45}r_{7-5}}
	\; t'_{56}
	\frac{\textrm{e}^{-\textrm{i}k_{1}c}}{\textrm{e}^{-2\textrm{i}k_{1}c}
		-r'_{56}r_{67}} \;  t'_{67}\nonumber\\ 
	&+&t_{67} 
	\frac{\textrm{e}^{-\textrm{i}k_{1}c}}{\textrm{e}^{-2\textrm{i}k_{1}c}
		-r'_{56}r_{67}} \; t_{56} \; 
	\frac{\textrm{e}^{-\textrm{i}k_{2}a}}{\textrm{e}^{-2\textrm{i}k_{2}a}
		-r'_{45}r_{7-5}}
	t_{45} \;   
	\frac{\textrm{e}^{-\textrm{i}k_{1}b}}{\textrm{e}^{-2\textrm{i}k_{1}b}
		-r'_{34}r_{7-4}} \;  t_{34} \;  \frac{1}{\textrm{e}^{-2\textrm{i} 
			k_{2}a}-r'_{23}r_{7-3}}  \; r'_{23}  \nonumber\\
	& \quad\times&\; t'_{34}  
	\frac{\textrm{e}^{-\textrm{i}k_{1}b}}{\textrm{e}^{-2\textrm{i}k_{1}b}
		-r'_{34}r_{7-4}} \; t'_{45} \; 
	\frac{\textrm{e}^{-\textrm{i}k_{2}a}}{\textrm{e}^{-2\textrm{i}k_{2}a}
		-r'_{45}r_{7-5}}
	 t'_{56} \;  
	\frac{\textrm{e}^{-\textrm{i}k_{1}c}}{\textrm{e}^{-2\textrm{i}k_{1}c}
		-r'_{56}r_{67}}  \; t'_{67}\\ 
	&+&t_{67} \; 
	\frac{\textrm{e}^{-\textrm{i}k_{1}c}}{\textrm{e}^{-2\textrm{i}k_{1}c}
		-r'_{56}r_{67}} \; t_{56} \; 
	\frac{\textrm{e}^{-\textrm{i}k_{2}a}}{\textrm{e}^{-2\textrm{i}k_{2}a}
		-r'_{45} r_{7-5}}
	\; t_{45} \;  
	\frac{\textrm{e}^{-\textrm{i}k_{1}b}}{\textrm{e}^{-2\textrm{i}k_{1}b}
		-r'_{34}r_{7-4}}  \; t_{34} \;  
	\frac{\textrm{e}^{-\textrm{i}k_{2}a}}{\textrm{e}^{-2\textrm{i} 
			k_{2}a}-r'_{23}r_{7-3}}  \nonumber\\
	& \quad\times&\; t_{23} \;  
	\frac{1}{\textrm{e}^{-2\textrm{i}k_{1}c}-r'_{12}r_{7-2}} \; r'_{12} \;	t'_{23} \; 
	\frac{\textrm{e}^{-\textrm{i}k_{2}a}}{\textrm{e}^{-2\textrm{i}k_{2}a}
		-r'_{23}r_{7-3}}
	\; t'_{34} \;  
	\frac{\textrm{e}^{-\textrm{i}k_{1}b}}{\textrm{e}^{-2\textrm{i}k_{1}b}
		-r'_{34}r_{7-4}} \; t'_{45} \;
	\frac{\textrm{e}^{-\textrm{i}k_{2}a}}{\textrm{e}^{-2\textrm{i}k_{2}a}
		-r'_{45}r_{7-5}}
	  t'_{56} \;  
	\frac{\textrm{e}^{-\textrm{i}k_{1}c}}{\textrm{e}^{-2\textrm{i}k_{1}c}
		-r'_{56}r_{67}}  \; t'_{67}\nonumber,
	\label{ec2:reflexion_primada}
\end{eqnarray}
and
\begin{equation}
	t'=t'_{12}
	\frac{\textrm{e}^{-\textrm{i}k_{1}c}}{\textrm{e}^{-2\textrm{i}k_{1}c}
		-r'_{12} r_{7-2}} \; t'_{23} 
	\frac{\textrm{e}^{-\textrm{i}k_{2}a}}{\textrm{e}^{-2\textrm{i}k_{2}a}
		-r'_{23} r_{7-3}}
	\;t'_{34}
	\frac{\textrm{e}^{-\textrm{i}k_{1}b}}{\textrm{e}^{-2\textrm{i}k_{1}b}
		-r'_{34} r_{7-4}} \;t'_{45}
	\frac{\textrm{e}^{-\textrm{i}k_{2}a}}{\textrm{e}^{-2\textrm{i}k_{2}a}
		-r'_{45} \; r_{7-5}}
	t'_{56}
	\frac{\textrm{e}^{-\textrm{i}k_{1}c}}{\textrm{e}^{-2\textrm{i}k_{1}c}
		-r'_{56} \; r_{67}} \; t'_{67}.
	\label{ec2:transmision_primada}
\end{equation}
In former equations we have used two different notations explained at the Appendix. The Hermitian case is recovered when $\gamma=0$, therefore $r=r'$ and $t=t'$ since the system has specular reflection and time reversal symmetry. We must emphasize that these equations are analytical and general expressions regardless of the nature of the waves as long as the definition of the scattering matrix for one-dimensional cases must be fulfilled~\cite{Robledo2020}.

\begin{figure}[h]
	\centering
	\includegraphics[width=8.6cm]{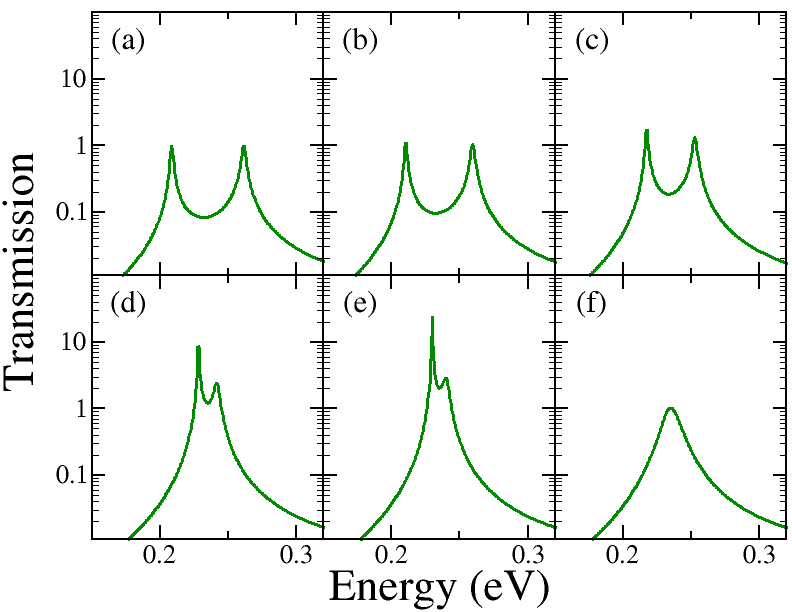}
	\caption{(Color online) Transmission spectra as a function of energy in which only the first doublet of the dimer is plotted. (a) For the Hermitian case ($\Gamma=0$) the maximum values of the resonances are found at $E_{1,1}=0.2086~\text{eV}$ (the first subscript refers to the associated doublet while the second one refers to the associated resonance), and $E_{1,2}=0.2615~\text{eV}$. Dimer parameters, in this and subsequent figures, are $a=1.15~$nm, $b=2c=0.02~\text{nm}$, $V_\textrm{b}= 50~\text{eV}$, and $V'=0~\text{eV}$. For $\Gamma=0$ the resonances have a unit amplitude and a maximum separation. As we increase the value of $\Gamma$ the separation between the resonances decreases and their amplitude increases before the EP as it is observed for $\Gamma$ equal to (b) 0.010, (c) 0.020. (d) 0.026 and (e) 0.0264. For $\Gamma=0.0264$ the resonances are already very close and overlap, losing the higher energy resonance. After the EP ($\Gamma\approx0.0266$) the width of the resonance increases and its amplitude decreases as observed in panel (f). The same values of $\Gamma$ are used in Figures~\ref{fig3:ReWavefunctions}, \ref{fig4:ImWavefunctions}, \ref{fig7:Phase}, \ref{fig8:Argand}, and \ref{fig9:Histogram}.}
	\label{fig2:TransmissionSpectrum}
\end{figure}
In Figure~\ref{fig2:TransmissionSpectrum} the first doublet of the dimer resonance spectrum is plotted as a function of energy. The resonances positions are obtained from the maxima of the total transmission ($T=|t|^2$). The Hermitian case is recovered for $\Gamma=0$ (see Figure~\ref{fig2:TransmissionSpectrum}(a)), which is related with $\gamma$ through
\begin{eqnarray}
\Gamma=\sqrt{\left[\frac{\hbar^2}{m}\gamma^2+(E-V')\right]^2-(E-V')^2}, \nonumber
\end{eqnarray} 
and the real and imaginary parts of the wavefunctions of these resonances are plotted in panel (a) of Figures~\ref{fig3:ReWavefunctions}~and~\ref{fig4:ImWavefunctions}, respectively. All the next figures are obtained when incoming waves are only from the left, {\it i.e.} $a_7=0$. As can be anticipated, the real part of the wavefunction associated to the lowest energy resonance of the doublet is a symmetric one, while the highest energy resonance wavefunction is antisymmetric. On the one hand, as we increase the $\Gamma$ value, the separation between the resonances decreases until they coalesce, as well as the width of the resulting resonance increases (see panels (b), (c), (d) and (e) of Figure~\ref{fig2:TransmissionSpectrum}). Also, the amplitude of the resonances increases but after passing the EP it starts decreasing its height (see Figure~\ref{fig2:TransmissionSpectrum}~(f)). This is outstanding because the amplitude of the resulting resonance is expected to grow indefinitely. On the other hand, as we increase $\Gamma$, the amplitude of the real and imaginary part of the wavefunction also increases. As can be seen in panels (b)-(e) of Figures~\ref{fig3:ReWavefunctions}~and~\ref{fig4:ImWavefunctions}, the amplitudes of the symmetric and antisymmetric modes grow although that of the symmetric mode grows more than the antisymmetric one. After passing the EP, in addition to the broken $\mathcal{PT}$-symmetry, both modes have coalesced into a mode that is neither symmetric nor antisymmetric. In fact, both modes coexist simultaneously and one is the complex conjugate of the other one. Figure~\ref{fig5:EP} shows the behavior of the real and imaginary part of the energy as a function of $\Gamma$. The real part is associated with the position of the maxima of each resonance, while the imaginary part is associated with their width. The result given by the Tight Binding model predicts a decay in the separation between the eigenenergies of the system as a square root~\cite{Kato,Heiss}. In Figure~\ref{fig6:Fit} it is shown the fit of the difference of the resonances as a function of $\Gamma$. In purple solid line the separation between the real parts of the resonances is plotted; the yellow solid line is a pure square root while the green solid line is the best fit that was obtained from data showing a power of 0.47.
\begin{figure}[h]
	\centering
	\includegraphics[width=8.6cm]{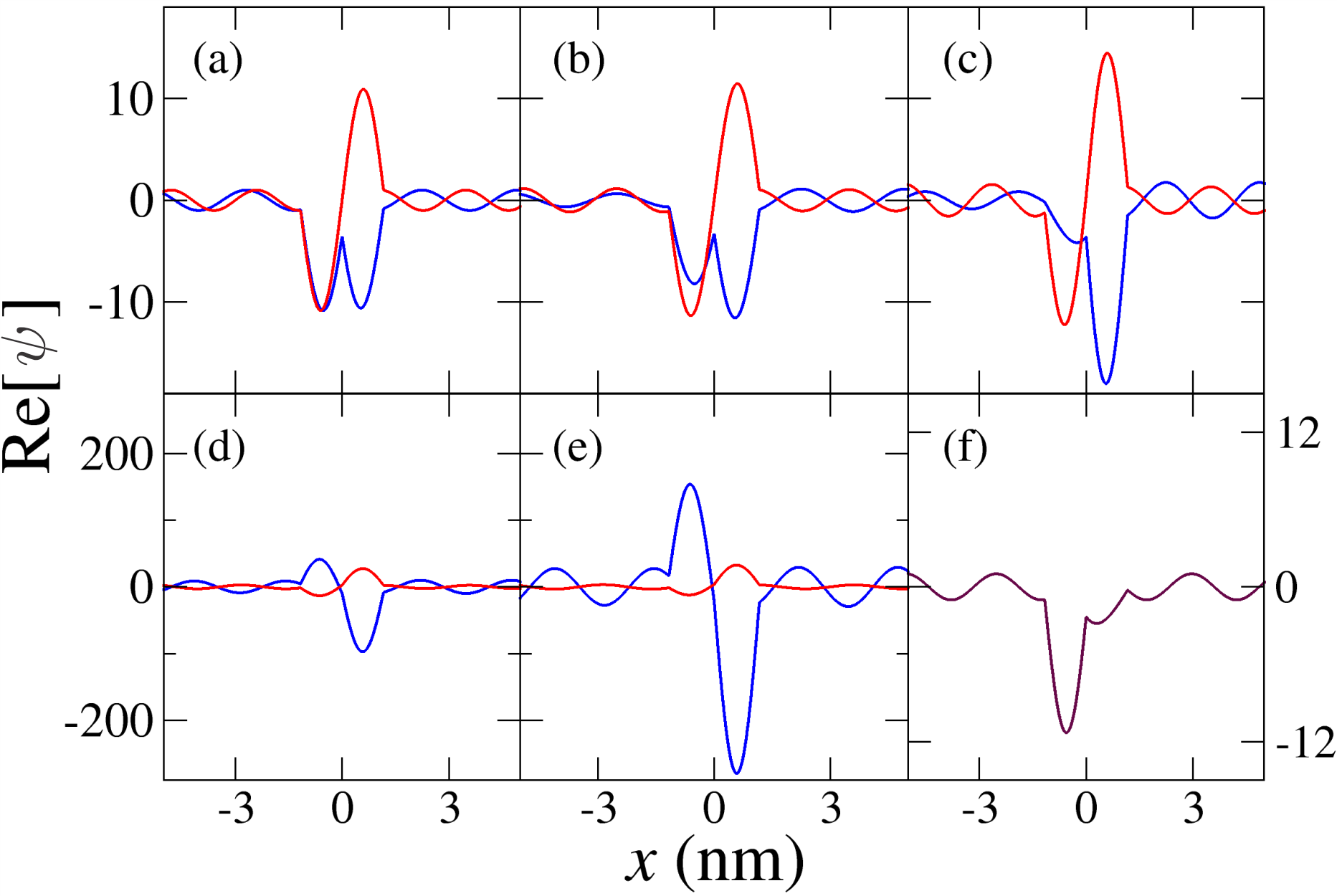}
	\caption{(Color online) Evolution of the real part (Re$[\psi]$) of the states associated to the resonances of the first doublet energy as the parameter $\Gamma$ is increased. The lowest energy resonance is plotted in blue while the one with highest energy is plotted in red, panels (a) to (e). The blue wavefunctions correspond to the symmetric state while the red ones are the antisymmetric states. Once the EP is reached only one resonance remains, therefore there is only one wave function as is shown in purple in panel (f). This mode is not symmetric nor antisymmetric. Before reaching the EP, the amplitude of the wave functions also increases until it decreases once the EP is passed.}
	\label{fig3:ReWavefunctions}
\end{figure}
\begin{figure}[h]
	\centering
	\includegraphics[width=8.6cm]{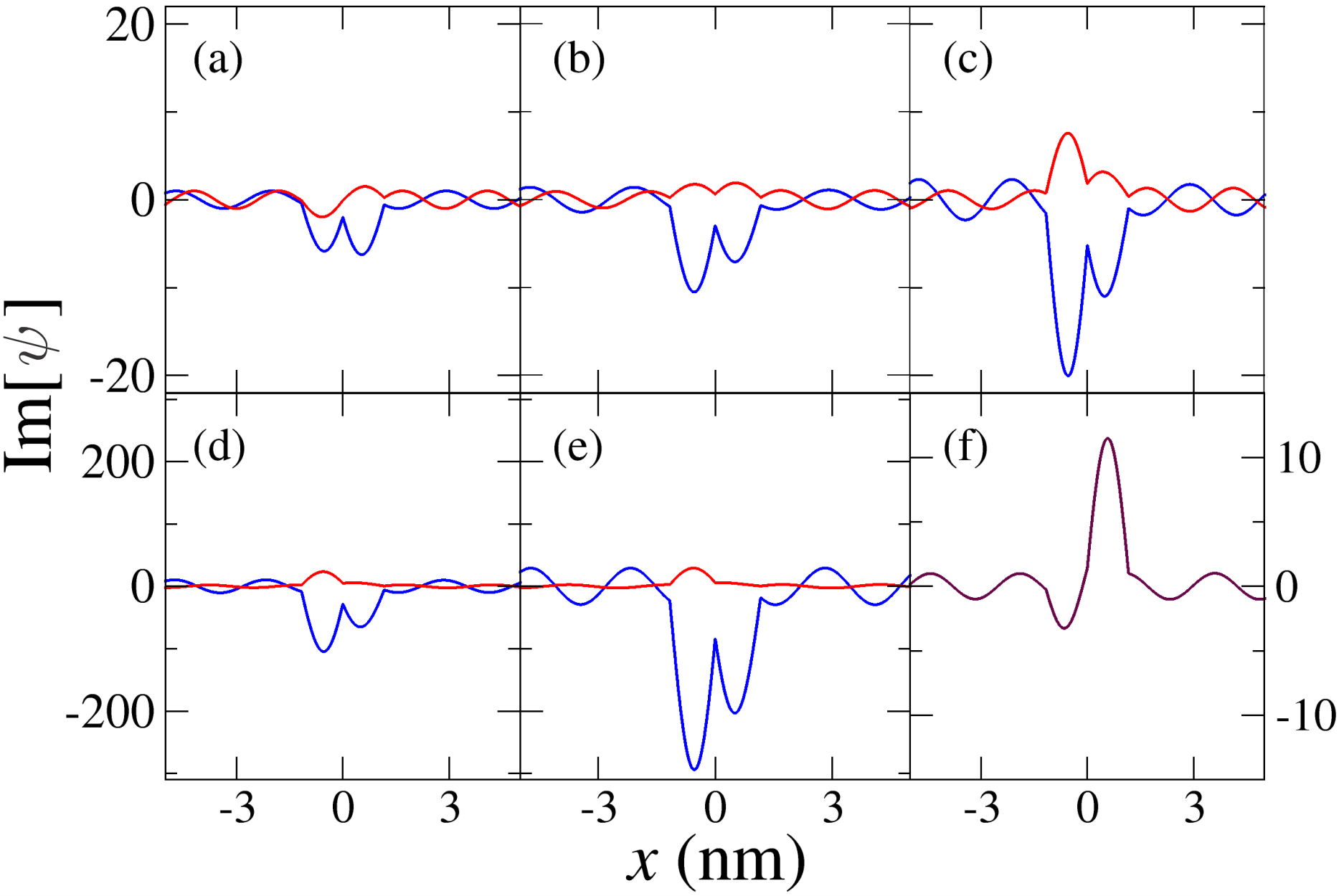}
	\caption{(Color online) Evolution of the imaginary parts of the wavefunctions (Im$[\psi]$) associated to the resonances of Figure~\ref{fig2:TransmissionSpectrum}. The blue, red, curve is the imaginary part of the lowest, highest, energy resonance. As in the real part, also the amplitudes increase as a function of $\Gamma$. Once the EP is passed there is only one resonance and the amplitude of the imaginary part of its wavefunction decreases as well as the real part.}
	\label{fig4:ImWavefunctions}
\end{figure}
%

%
\begin{figure}[h]
	\centering
	\includegraphics[width=8.6cm]{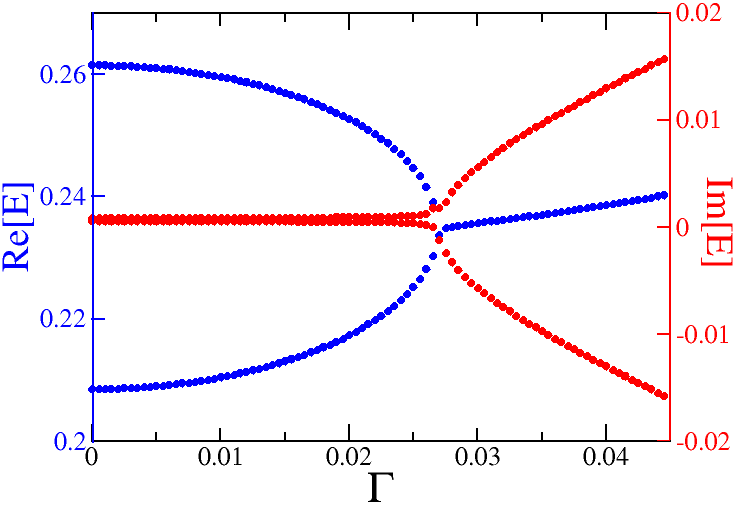}
	\caption{(Color online) EP of the dimmer. On the one hand, the real part corresponds to the positions of the resonances. For the Hermitian case, the separation of the resonances is given by the natural level repulsion. As $\Gamma$ increases, the separation between the resonances decreases until both coalesce at the EP. After the EP there is only one resonance since both have coalesced. On the other hand, the imaginary part corresponds to the widths of the resonances, measured by means of the $Q$ factor. As we get closer to the EP, the resonances overlap increasing their widths until they coalesce into a single one increasing significantly its width.}
	\label{fig5:EP}
\end{figure}
\begin{figure}
	\centering
	\includegraphics[width=8.6cm]{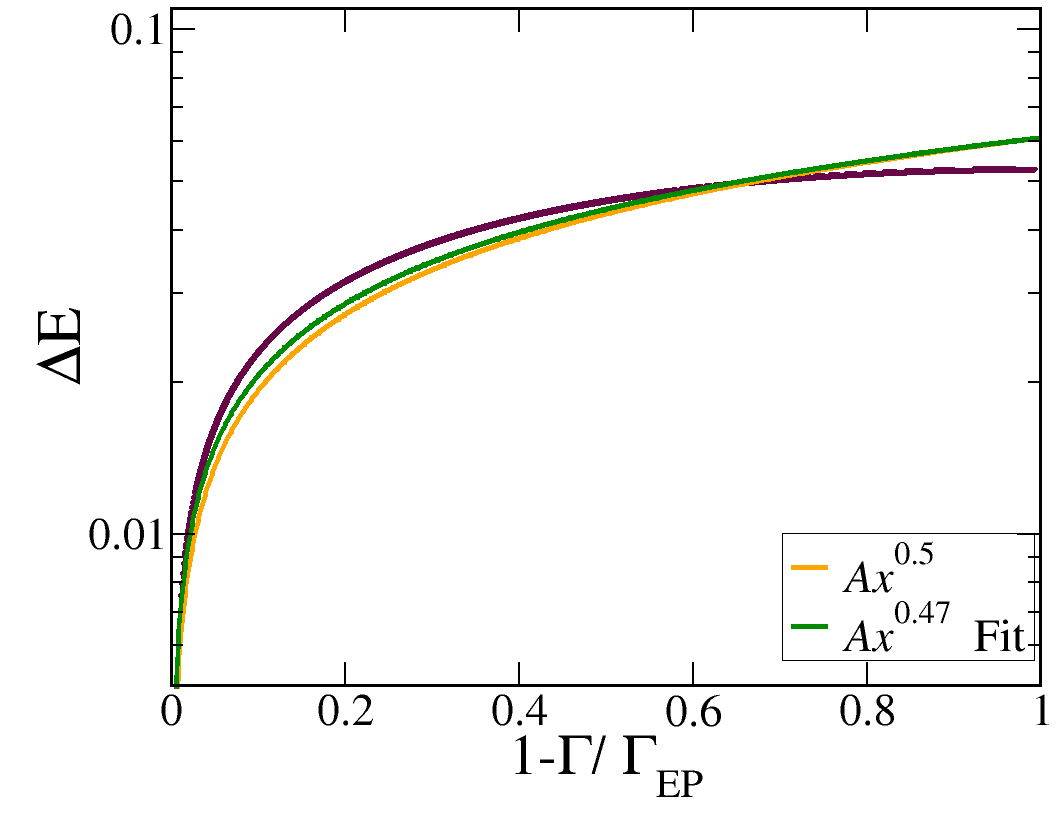}
	\caption{(Color online) The separation of the resonances as a function of $\Gamma$ is shown in purple. This is compared with the obtained fit, in green, which corresponds to the model $y=A x^{B}$, where $A=0.061$ and $B=0.47$, the latter being the power obtained that is close to the square root predicted by the tight binding model, plotted in yellow.}
\label{fig6:Fit}
\end{figure}


The scattering matrix formalism allows to us to study not only the transmission and reflection amplitudes but also the phases of the system. The eigenphases of the $S$ matrix, given after diagonalizing Eq.~\ref{ec2:Smatrix}, are
\begin{eqnarray}
	\begin{bmatrix}
		e^{\textrm{i} \theta} & 0\\
		0 & e^{\textrm{i} \theta'} 
	\end{bmatrix}=
	\begin{bmatrix}
		\frac{1}{2}\left( r+r'+\sqrt{(r-r')^2+4tt'}\right) & 0\\
		0 &    \frac{1}{2}\left( r+r'-\sqrt{(r-r')^2+4tt'}\right)
	\end{bmatrix} 
	\label{ec2:matriz_diagonalizada}.
\end{eqnarray}
For $\Gamma=0$ also the Hermitian eigenphases are recovered~\cite{Mares2009}. In general, the eigenphases are complex and can be written as
\begin{equation}
	\theta = \theta_{\text{Re}} + \textrm{i}  \; \theta_{\text{Im}}, \nonumber
\end{equation}
where $\theta_{\text{Re}}$ is the real part of the phase, while $\theta_{\text{Im}}$ is its imaginary part. Accordingly, when separating the real and imaginary part of $\textrm{e}^{\textrm{i}\theta}$ we have
\begin{equation}
	\textrm{e}^{\textrm{i}\theta}=     
	\textrm{e}^{\textrm{i}(\theta_{\text{Re}} + \textrm{i} \; 
		\theta_{\text{Im}})}= \textrm{e}^{\textrm{i} \theta_{\text{Re}}}  
	\textrm{e}^{-\theta_{\text{Im}} }  
	\label{eq3:repart_impart}.
\end{equation}
The former equation can be interpreted as a module multiplied by a complex number of unitary magnitude. In other words, $\textrm{e}^{\textrm{i} \theta_{\text{Re}}} $ is a unitary complex number that, by varying the energy $E$, forms a circle in the Argand plane, while $\textrm{e}^{-\theta_{\text{Im}} }$ acts as the module of the circle. 

\begin{figure}
	\centering
	\includegraphics[width=8.6cm]{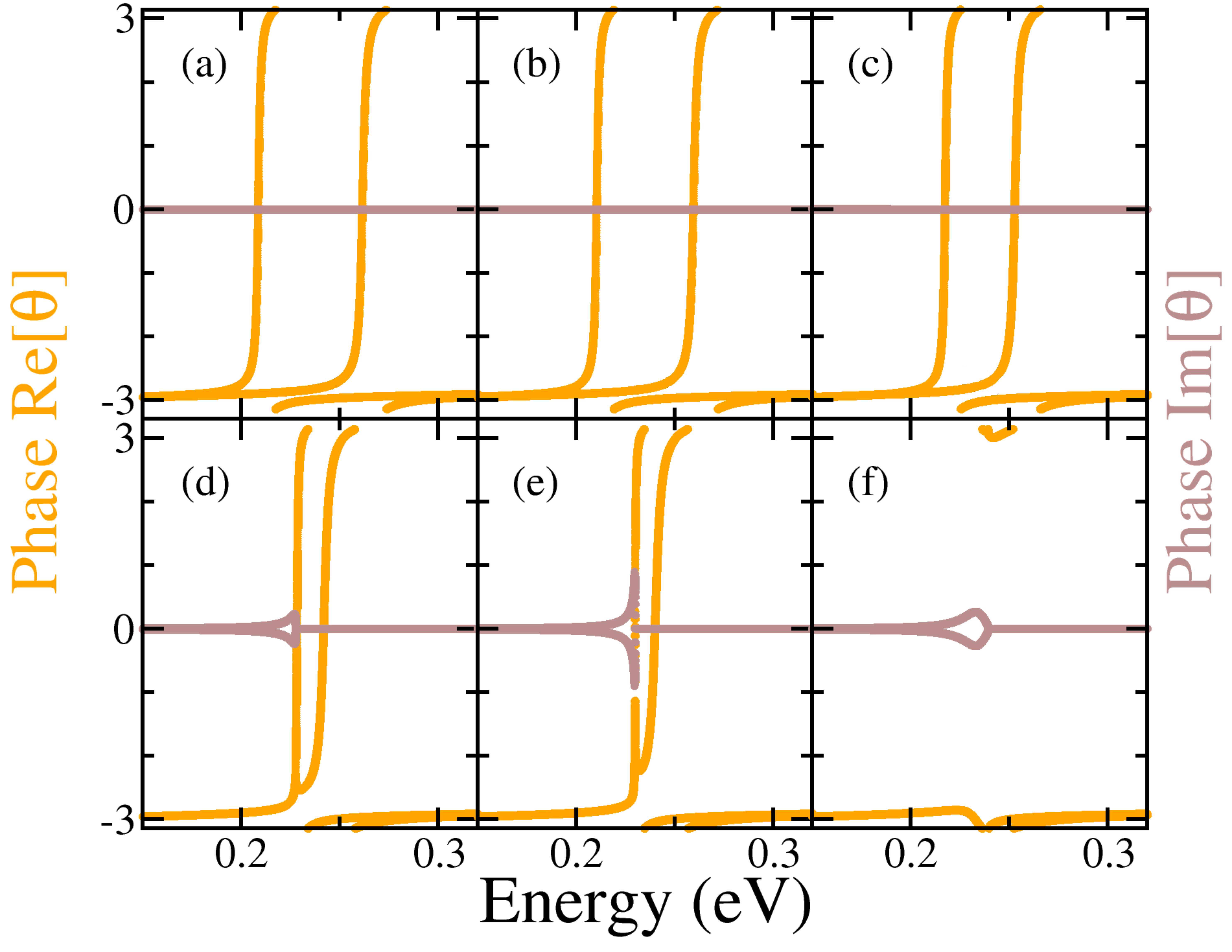}
	\caption{(Color online) Behaviour of the real (yellow solid line) and imaginary (brown solid line) parts of the phases $\theta$ and $\theta'$ as a function of the energy. (a) In the Hermitian case the separation of the resonances is maximum and the imaginary part of the phase is zero. As $\Gamma$ grows the separation of the resonances diminishes (panels (b) and (c)). (d) When we approach to the EP the imaginary part stops being null and a protuberance appears near the lowest energy resonance. (e) The closer we get to the EP the more the protuberance grows. (f) Once we past the EP, the real part of the phases lives only in the neighborhood of $-\pi$ and $\pi$, and the protuberance becomes smaller.}
	\label{fig7:Phase}
\end{figure}

In Figure~\ref{fig7:Phase} we plot the real and the imaginary part of the phase as a function of the energy. Panel (a) corresponds to the Hermitian case in wich the real part has two separated resonances that correspond to $\theta$ and $\theta'$. In this case the imaginary part is null. As $\Gamma$ increases the distance between resonances becomes smaller but the imaginary part still been zero as can be seen in panels (b) and (c). In panel (d) the resonances overlap and the imaginary part becomes not null for positive and negative values. This protuberance appears near the lowest energy resonance, which is the one that prevails. As we approach to the EP the separation of the resonances reduces while the protuberance grows (panel (e)). Once the EP value has been exceeded the real part of the phase is restricted to values close to $-\pi$ and $\pi$ and the imaginary part remains not null in the neighborhood of the remain resonance as is plotted in panel (f).

\begin{figure}
	\centering
	\includegraphics[width=8.6cm]{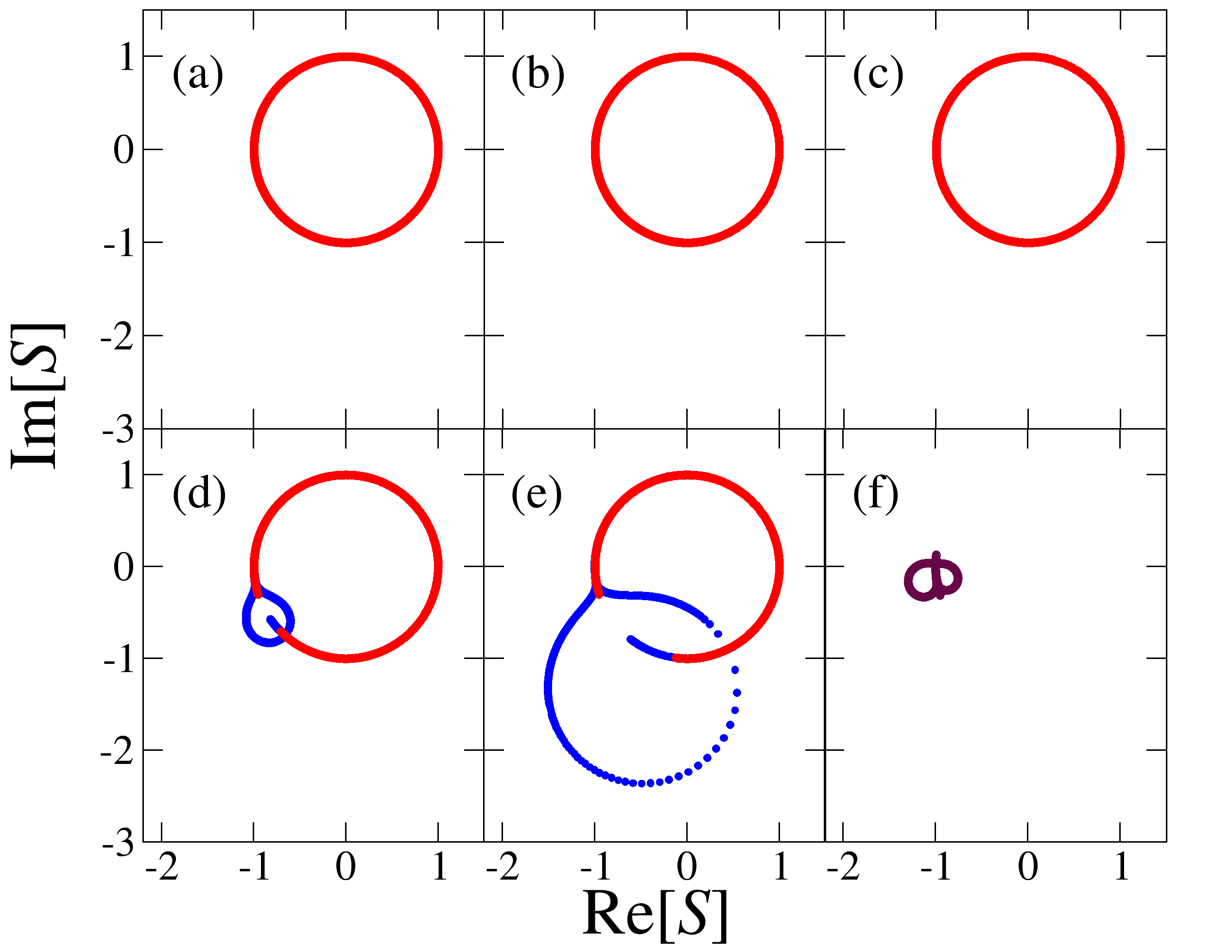}
	\caption{(Color online) Phase distribution of both resonances in Argand plane. Blue points correspond to the lowest energy resonance while the red ones correspond to the highest energy resonance. In (a), (b), and (c) all the points live in the unit circle because $\theta_{\text{Im}}=0$ (blue points are below the red ones). In panels (d) and (e) the red resonance still lives in the unit circle but a lobe appears in the lowest energy resonance. This lobe increases its size up to infinity where it changes its concavity to reduce its size as $\Gamma$ keeps growing. In panel (f) the coalesced resonance has diminished its size and distribution.}
	\label{fig8:Argand}
\end{figure}

Figure~\ref{fig8:Argand} shows the phase of both resonances in Argand plane. Blue points are the lowest energy resonance while red points are the highest energy resonance. In panels (a), (b), and (c) all of the points of each resonance are distributed in the unit circle since the imaginary part is equal to zero. In panel (d) the protuberance of the imaginary part makes that a lobe appears in the lowest energy resonance whereas that the highest energy resonance stops filling completely the circle. As the $\Gamma$ value is increased even more, the lobe increases its size as does the empty space of the resonance (see panel (e)). In fact, the size of the lobe increases up to infinity and changes its concavity past the EP to reduce its size again. For its part, the distribution of the phase of the other resonance ceases to be distributed over the entire circle until only a small arc remains as can be seen in panel (f) where only one resonance remains.

\begin{figure}
	\centering
	\includegraphics[width=8.6cm]{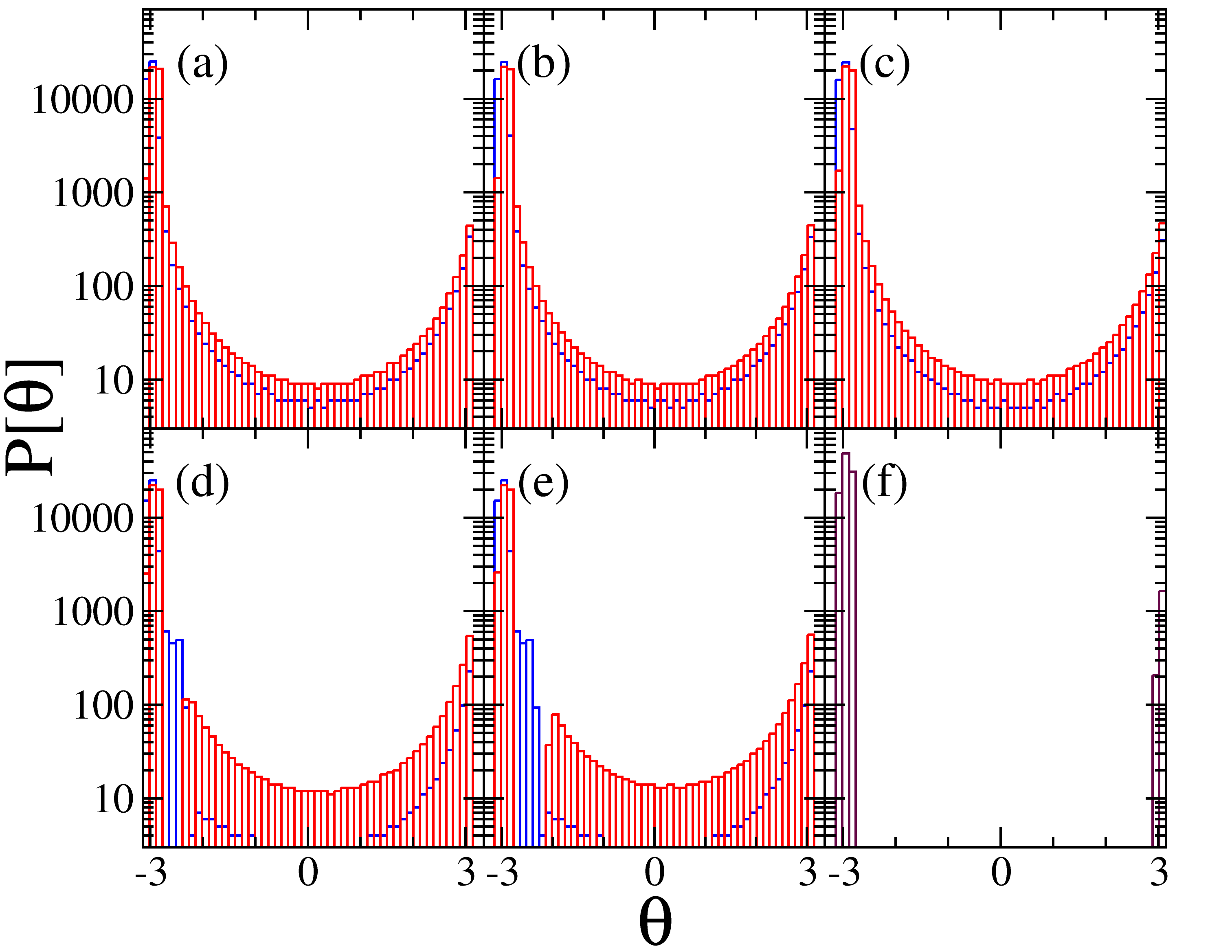}
	\caption{(Color online) Histogram of the phase distribution of the resonances as a function of $\Gamma$. Blue (red) histogram corresponds to the lowest (highest) energy resonance, except for the coalesced resonance plotted in purple. As can be inferred from the Argand plane, the phases are distributed all over the angles but not in a uniform way as is shown in (a), (b) and (c). When the lobe appears the resonances stop visiting all of the angles hence empty spaces appear in the distribution (panels (d) and (e)). In (f) all of the points are distributed only around $-\pi$ and $\pi$.}
	\label{fig9:Histogram}
\end{figure}

Finally, in Figure~\ref{fig9:Histogram} the histogram of the phase distribution of the two resonances under study between $-\pi$ and $\pi$ is shown. The histogram in blue corresponds to the resonance with the lowest energy, also in blue, while the histogram in red corresponds to the resonance with the highest energy, also in red (see Figure~\ref{fig8:Argand}). When the phase is distributed over the circle, in the Argand plane, its distribution takes values in the entire range from $-\pi$ to $\pi$ (panels (a), (b), and (c)). When the lobe appears, as well as the empty space, this distribution changes starting to create holes (panels (d) and (e)). These gaps in the distributions grow until the histograms overlap into only one (the resonances have already coalesced) resulting in a distribution of points close to the neighborhoods of the $-\pi$ and $\pi$ values shown in panel (f) (purple).

\section{Concluding remarks}

We have shown that the amplitude of the resonances of a ${\cal PT}$-symmetric system which presents an EP grows as $\Gamma$ get closer to the transition value. At the EP the resonances coalesce in energy as well as the eigenstates. The amplitude of the remain resonance starts decreasing as $\Gamma$ keeps growing. The eigenphases show the same behaviour by increasing their radii up to infinity and becoming smaller once we have past the EP. The distribution of the phase changes dramatically starting from being distributed in all over the angles to be localized in the vicinity of a well defined region. Since we have developed general expressions for the scattering matrix these behaviours can be tested experimentally in microwaves, or elastic, regime by building a proper system with slabs, or notches, outside the gain and loss regions in the one-dimensional approximation. Our desire that the results presented in this letter will contribute to the generalization of a non-Hermitian quantum theory.

\begin{acknowledgments}
	J.~C.-G. thanks financial support from CONAHCyT. V.~D.-R is grateful
	with the Sistema Nacional de Investigadores, Mexico, and with E. Michel-Hern\'andez and G.~T. Dom\'inguez-Michel for their
	encouragement. G.~B. and V.~D.-R. thank financial support for the CONAHCyT project CB2017-2018/A1-S-33920 and the UAM-Azc project CB004-22. The authors thank useful comments and discussions of J. A. Franco-Villafa\~ne, A. A. Fern\'andez-Mar\'in and M. Mart\'inez-Mares.
	
\end{acknowledgments}

\section{Appendix}
\subsection{$S$ matrix building process}

The first notation uses two joined subscripts. For example, in $r'_{\eta \delta}$ we refer to the reflection that occurs between the adjacent potentials $\eta$ and $\delta $ when waves incide from the right. The scattering matrix for this case is given by
\begin{equation}
	S_{\delta \eta}=\begin{bmatrix}
		r_{\delta \eta}  & t'_{\delta \eta} \\
		t_{\delta \eta}     & r'_{\delta \eta}
	\end{bmatrix}
	=\begin{bmatrix}
		\frac{k_{\delta}-k_{\eta}}{k_{\delta}+k_{\eta}}  &  
		\frac{2k_{\eta}}{k_{\delta}+k_{\eta}}\\
		\frac{2k_{\delta}}{k_{\delta}+k_{\eta}}    & 
		-\frac{k_{\delta}-k_{\eta}}{k_{\delta}+k_{\eta}}
	\end{bmatrix}  \label{ec2:matriz_scattering_delta_eta},
\end{equation}
where $k_\delta$ and $k_\eta$ are the wave numbers of the region $\delta$ and $\eta$, respectively, while $S_{\delta\eta}$ is the scattering matrix between the $\delta$ and $\eta$ regions which also represents the scatterer between those regions.

The second notation uses two subscripts separated by a hyphen. Combinating the scattering matrices $S_{\delta\eta}$ and $S_{\eta\omega}$ gives $S_{\delta-\omega}$. For example, $t_{\delta-\omega}$ refers to the transmission amplitude generated when the waves are transmitted from $\delta$ region to the non-contiguous $\omega$ region when incident waves arrive from the left. Part of the incident wave from the left on the scatterer $S_{\delta \eta}$ is reflected back into the $\delta$ region and the other part is transmitted towards the $\eta$ region. The part that was transmitted travels to the scatterer $S_{\eta\omega}$, gaining a phase in this path, in which a fraction of that wave is reflected in the $\eta$ region and the other is transmitted towards the $\omega$ region. The wave that was reflected travels again to the scatterer $S_{\delta\eta}$ in which a part of that wave is reflected again into the $\eta$ region and the other part is transmitted to the $\delta$ region. This process of multiple reflections in the region $\eta$ is repeated infinitely. On the one hand, the sum of all the waves in the $\omega$ region is a geometric series which results in $t_{\delta-\omega}$. On the other hand, the sum of the waves in the $\delta$ region, which is also a geometric series, gives $r_{\delta-\omega}$. These two elements are the transmission and reflection amplitudes when waves are incident from the left of the matrix
\begin{eqnarray}
	&S&_{\delta-\omega}=\begin{bmatrix}
		r_{\delta-\omega}	& t'_{\delta-\omega}   \\
		t_{\delta-\omega}	&  r'_{\delta-\omega}
	\end{bmatrix}\nonumber\\
	&=&\begin{bmatrix}
		r_{\delta\eta} +t'_{\delta 
			\eta}\frac{1}{\textrm{e}^{-2\textrm{i} k_{\eta} d } -r'_{\delta 
				\eta}r_{\eta \omega}}r_{\eta \omega}t_{\delta \eta} & t'_{\delta 
			\eta}\frac{\textrm{e}^{-\textrm{i} k_{\eta} d 
		}}{\textrm{e}^{-2\textrm{i} k_{\eta} d } -r'_{\delta \eta}r_{\eta 
				\omega}}t'_{\eta \omega} \\
		t_{\eta \omega}\frac{\textrm{e}^{-\textrm{i} k_{\eta} d 
		}}{\textrm{e}^{-2\textrm{i} k_{\eta} d } -r'_{\delta \eta}r_{\eta 
				\omega}}t_{\delta \eta} & r'_{\eta \omega} +t_{\eta 
			\omega}\frac{1}{\textrm{e}^{-2\textrm{i} k_{\eta} d } -r'_{\delta 
				\eta}r_{\eta \omega}}r'_{\delta \eta}t'_{\eta \omega} 
	\end{bmatrix}.\nonumber
	\label{ec2:combinacion matrices}
\end{eqnarray}
This procedure is also valid if it is necessary to build the system from the right since $S=S_{1-7}=S_{7-1}$. In particular, this combination procedure is performed five times for the dimer by adding potential barriers from left to right. A part of the analytical expression of the reflection amplitude (Equation~2) of the dimer is
\begin{eqnarray}
		r&=& r_{12}   \nonumber\\
		&+&t'_{12}\; \frac{1}{\textrm{e}^{-2\textrm{i} 
				k_1c}-r'_{12}\;r_{23}} \; r_{23}  \;t_{12} \nonumber\\
		&+&t'_{12}\; 
		\frac{\textrm{e}^{-\textrm{i}k_{1}c}}{e^{-2\textrm{i} 
				k_1c}-r'_{12}\;r_{23}}\;  t'_{23} 
		\;\frac{1}{\textrm{e}^{-2\textrm{i}k_{2}a}-r'_{1-3}\;r_{34}}\;r_{34}\;t_{23}\; \frac{\textrm{e}^{-\textrm{i} k_1 
				c}}{\textrm{e}^{-2\textrm{i} k_1c}-r'_{12}\;r_{23}} \;t_{12}+\cdots.
       \label{eqapp1:reflection}
\end{eqnarray}
Here $r_{12}$ is the first reflection that occurs when waves are incident from the left on the boundary between regions I and II. The second term of $r$ represents the wave that was transmitted from region I to region II, it suffers a process of multiple reflections between regions II and III, and is transmitted from region II to region I. The third term represents the wave that is transmitted from region I to region II, it suffers a process of multiple reflections between the boundaries of region II, is transmitted from region II to region III, is reflected multiple times between the boundaries of region I and region IV (note the term $r'_{1-3}$ in the denominator), is transmitted from region III to region II, reflected multiple times in region II, and is finally transmitted from region II to region I. The physical interpretation continues in the same way for subsequent terms. The interpretation of the transmission amplitude follows an equivalent process to that of $r$.

\end{document}